\documentclass[10pt,letterpaper,twocolumn]{article} %

\usepackage{ol2}
\usepackage[draft]{hyperref}
\usepackage{amsmath}
\usepackage{amssymb}
\usepackage{comment}
\usepackage{siunitx}
\usepackage{color}

\usepackage{epstopdf}

\def\jap{J.\ Appl.\ Phys.\ }

\allowdisplaybreaks[1]

\begin{document}

\twocolumn[ 

\title{Wavelength conversion and parametric amplification of optical pulses via quasi-phase-matched FWM in long-period Bragg silicon waveguides}


\author{S. Lavdas,$^{1}$ S. Zhao,$^1$ J. B. Driscoll,$^2$ R. R. Grote,$^2$ R. M. Osgood, Jr.,$^2$ N. C. Panoiu$^{1,*}$}

\address{
$^1$Department of Electronic and Electrical Engineering, University College London, Torrington
Place, London WC1E, UK
\\
$^2$Microelectronics Sciences Laboratories, Columbia University,New York, NY 10027,  USA

$^*$Corresponding author: n.panoiu@ucl.ac.uk}

\begin{abstract}We present a theoretical analysis supported by comprehensive numerical simulations
of quasi phase-matched four-wave mixing (FWM) of ultrashort optical pulses that propagate in
weakly width-modulated silicon photonic nanowire gratings. Our study reveals that, by properly
designing the optical waveguide such that the interacting pulses co-propagate with the same
group-velocity, a conversion efficiency enhancement of more than \SI{15}{\decibel}, as compared to
a uniform waveguide, can readily be achieved. We also analyze the dependence of the conversion
efficiency and FWM gain on the pulse width, time delay, walk-off parameter, and grating modulation
depth.
\end{abstract}


] 

\noindent Frequency generation in optical systems is the main underlying process in a series of
key applications, including all-optical signal processing, optical amplification, and wavelength
multiplexing. One of the most facile approaches to achieve this functionality is via optical-wave
interaction in nonlinear media. In the case of media with cubic nonlinearity, the simplest such
interaction is four-wave mixing (FWM), a nonlinear process in which two photons combine and
generate a pair of photons with different frequencies. Due to its simplicity and effectiveness,
FWM has been at the center of intense research, from the early days of nonlinear fiber optics
\cite{sba74apl,hjk78jap} to the recent studies of FWM in ultra-compact silicon (Si) devices
\cite{crd03oe,drc04oe,fys05oe,edo05oe,fts06n,lzf06oe,pco06ol,yft06ptl,krs06oe,fts07oe,lov10np,zpm10np,klo11oe,lkr12np,dog12oe,lpa07oe,opd09aop}.
Silicon photonic nanowire waveguides (Si-PNWs) are particularly suited to achieve highly efficient
FWM, as Si has extremely large cubic nonlinearity over a broad frequency domain. Equally important
in this context, due to the deep-subwavelength size of the cross-section of Si-PNWs, the
parameters quantifying their optical properties depend strongly on wavelength and waveguide size
\cite{lpa07oe,opd09aop}. As a result, one can easily control the strength and phase-matching of
the FWM. These ideas have inspired intense research in chip-scale devices based on FWM in Si
waveguides, with optical parametric amplifiers \cite{fts06n,lzf06oe,lov10np}, frequency converters
\cite{yft06ptl,krs06oe,fts07oe,zpm10np,klo11oe,lkr12np,dog12oe}, sources of quantum-correlated
photon pairs \cite{la06ol}, and optical signal regenerators \cite{sft08np} being demonstrated.

One of the main properties of Si-PNWs, which makes them particularly suitable to achieve efficient
FWM, is that by properly designing the waveguide geometry one can easily engineer the dispersion
to be either normal or anomalous within specific spectral domains. More specifically, Si-PNWs with
relatively large cross-section have normal dispersion, which precludes phase matching of the FWM.
This drawback can be circumvented by scaling down the waveguide size to a few hundred of
nanometers as then the dispersion becomes anomalous. The price one pays for this small
cross-section is that the device operates at reduced optical power. An alternate promising
approach to achieve phase-matched FWM in the normal dispersion regime is to employ
quasi-phase-matching (QPM) techniques, i.e. to cancel the linear and nonlinear phase mismatch of
the interacting waves by periodically varying the waveguide cross-section. This technique has been
recently used for \textit{cw} optical beams \cite{dog12oe}, yet in many cases of practical
importance it is desirable to achieve FWM in the pulsed regime. In addition, at large power
\textit{cw} beams are strongly depleted by optical losses, which results in the detuning of the
FWM.

In this Letter we show that efficient QPM FWM of optical pulses can be achieved in Si-PNWs whose
width varies periodically along the waveguide. In this work we focus on the QPM FWM of pulses that
propagate in the normal dispersion regime, as in this case one cannot apply alternative phase
shifting methods based on nonlinearly induced phase-shifts. Our analysis of the FWM in long-period
Bragg Si-PNWs is based on a theoretical model introduced in \cite{cpo06jqe}, which fully describes
optical pulse propagation and the influence of free-carriers (FCs) on the optical field dynamics
(see also \cite{dog12oe,plo09ol,ldj13ol}):
\begin{subequations}
\label{tm}
\begin{align}
\label{uzt}i\frac{\displaystyle \partial u}{\displaystyle \partial z}&+\sum\limits_{n= 1}^{n=4}
\frac{\displaystyle i^{n}\beta_{n}(z)}{\displaystyle n!}\frac{\displaystyle \partial^{n}
u}{\displaystyle \partial t^{n}}= -i\left[\frac{\displaystyle c\kappa(z)}{\displaystyle 2nv_{g}(z)}\alpha_{\mathrm{fc}}(z)+\alpha\right]u \nonumber \\
-&\frac{\displaystyle \omega \kappa(z)}{\displaystyle n v_{g}(z)}\delta n_{\mathrm{fc}}(z)u
-\gamma(z)\left[1+i\tau(z)\frac{\partial}{\partial t}\right]\vert u\vert^{2} u, \\
\label{dens}\frac{\displaystyle \partial N}{\displaystyle \partial t} &= -\frac{\displaystyle
N}{\displaystyle t_{c}} + \frac{\displaystyle 3 \Gamma^{\prime\prime}(z)}{\displaystyle
4\epsilon_{0}\hbar A^{2}(z)v_{g}^{2}(z)} |u|^{4},
\end{align}
\end{subequations}
where $u(z,t)$ and $N(z,t)$ are the pulse envelope and FC density, respectively, $t$ is the time,
$z$ is the distance along the waveguide, $\beta_{n}=d^{n}\beta/d\omega^{n}$ is the $n$th order
dispersion coefficient, $\kappa(z)$ is the overlap between the optical mode and the (Si) active
area of the waveguide, $v_{g}(z)$ is the group-velocity, $\delta n_{\mathrm{fc}}(z)$
[$\alpha_{\mathrm{fc}}(z)$] are $N$-dependent FC-induced index change (losses) \cite{sb87jqe}, and
$\alpha$ is the waveguide loss ($\alpha=0$ unless otherwise stated). The nonlinear coefficient,
$\gamma$, is given by $\gamma(z)=3\omega \Gamma(z)/4\epsilon_{0}A(z)v_{g}^{2}(z)$, and the shock
time scale is $\tau(z)=\partial\ln \gamma(z)/\partial \omega$, where $A(z)$ and $\Gamma(z)$ are
the cross-sectional area and the effective third-order susceptibility, respectively. The system
\eqref{tm} is integrated numerically by using a split-step Fourier method \cite{opd09aop}. Also,
in this study we set $t_{c}=$~\SI{1}{\nano\second}.
\begin{figure}[t]
\centerline{\includegraphics[width=8cm]{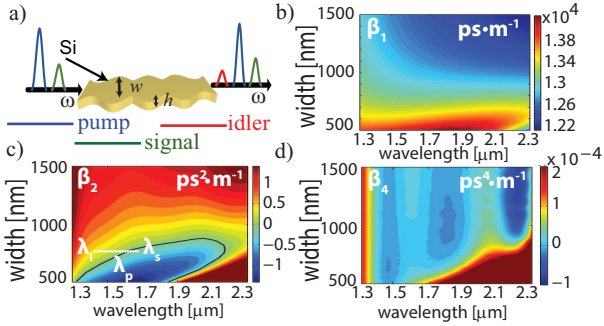}} \caption{(a) Schematics showing a
periodically width-modulated Si-PNW and the configuration of a pulsed seeded degenerate FWM
set-up. Dispersion maps of dispersion coefficients: (b) $\beta_1$, (c) $\beta_2$, and (d)
$\beta_4$.} \label{geom_prop}
\end{figure}

The optical waveguide considered here consists of a Si core with constant height,
$h=$~\SI{250}{\nano\meter}, and periodically modulated width, $w(z)$, buried in
$\mathrm{SiO_{2}}$. We assume a sinusoidal dependence, $w(z)=w_0+\Delta w \sin(2\pi z/\Lambda)$,
where $w_{0}$, $\Delta w$, and $\Lambda$ are the average width, amplitude of the width modulation,
and its period, respectively, but more intricate profiles $w(z)$ can be readily investigated by
our method. As illustrated in Fig.~\ref{geom_prop}(a), we consider the case of degenerate FWM, in
which two photons at the pump frequency, $\omega_{p}$, interact with the nonlinear medium and
generate a pair of photons at signal ($\omega_{s}$) and idler ($\omega_{i}$) frequencies. This FWM
process is most effective when
\begin{equation}
\left| 2(\beta_p-\gamma^{\prime} P_p)-\beta_s-\beta_i\right|=K_{g}, \label{Dbeta}
\end{equation}
where $K_{g}=2\pi/\Lambda$ is the Bragg wave vector, $P_{p}$ is the pump peak power, and
$\beta_{p,s,i}(\omega)$ are the mode propagation constants evaluated at the frequencies of the
co-propagating pulses. Note that in Eq.~\eqref{Dbeta} all width-dependent quantities are evaluated
at $w=w_{0}$.

If $\Delta\omega\equiv\omega_{s}-\omega_{p}=\omega_{p}-\omega_{i}\ll\omega_{p}$, Eq.~\eqref{Dbeta}
can be cast to a form that makes it more suitable to find the wavelengths of the
quasi-phase-matched pulses by expanding in Taylor series the functions $\beta_{p,s,i}(\omega)$,
around $\omega_{p}$. Keeping the terms up to the fourth-order, Eq.~\eqref{Dbeta} becomes:
\begin{equation}\label{DbetaTaylor}
    \left\vert2\gamma^{\prime}
    P_p+\beta_{2,p}\Delta\omega^2+\beta_{4,p}\Delta\omega^4/12\right\vert=K_{g}.
\end{equation}

\begin{figure}[t]
\centerline{\includegraphics[width=8cm]{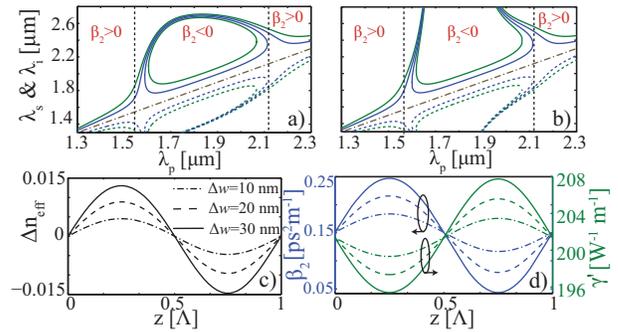}} \caption{(a), (b) Wavelength diagrams
defined by the phase-matching conditions \eqref{Dbeta} and \eqref{DbetaTaylor}, respectively.
Solid (dashed) lines correspond to the signal (idler) and green (blue) lines to
$\Lambda=$~\SI{2}{\milli\meter} ($\Lambda=$~\SI{6}{\milli\meter}). Dash-dot lines correspond to
$\lambda_{p}=\lambda_{s}=\lambda_{i}$ and vertical dotted lines mark $\beta_{2}(\lambda)=0$.
$z$-dependence of $\Delta n_{\mathrm{eff}}$, (c), and $\beta_2$, and $\gamma^{\prime}$, (d), shown
for one period, $\Lambda$. In (c) and (d) the lines correspond to $\Delta w=$~\SI{10}{\nano\meter}
(-$\cdot$-), $\Delta w=$~\SI{20}{\nano\meter} (- - -), and $\Delta w=$~\SI{30}{\nano\meter} (---).
In all panels $w_0=$~\SI{740}{\nano\meter}.} \label{Pha_match}
\end{figure}
The dispersive properties of the Si-PNW, summarized in Fig.~\ref{geom_prop}, define the spectral
domain, in which efficient FWM can be achieved. The width dependence of the dispersion
coefficients and other relevant waveguide parameters, i.e. $\gamma$, $\kappa$, and $\tau$, was
obtained by using a method described in detail in \cite{dog12oe,ldj13ol}. Importantly, with a
proper choice of the operating wavelength or waveguide width, the photonic wire can have both
normal and anomalous GVD. The wavelengths, for which the FWM is quasi-phase-matched and determined
from Eqs.~\eqref{Dbeta} and \eqref{DbetaTaylor}, are plotted in Figs.~\ref{Pha_match}(a) and
\ref{Pha_match}(b), respectively. These results show that, as expected, for relatively small
$\Delta\omega$, Eqs.~\eqref{Dbeta} and \eqref{DbetaTaylor} lead to similar predictions, whereas
they disagree for large $\Delta\omega$. Interestingly enough, Fig.~\ref{Pha_match}(a) shows that
for certain $\lambda_{p}$'s FWM can be achieved at more than one pair of wavelengths,
$(\lambda_{s},\lambda_{i})$, meaning that optical bistability could readily be observed in this
system. The corresponding $z$-dependence over one period of the variation of the effective modal
refractive index, $\Delta n_{\mathrm{eff}}$, $\beta_{2}$, and $\gamma^{\prime}$, is presented in
Figs.~\ref{Pha_match}(c) and \ref{Pha_match}(d).

The wavelength conversion efficiency and parametric amplification gain are determined from the
pulse spectrum. Thus, we launch into the waveguide pulses whose temporal profile,
$u(0,t)=\sqrt{P_p}[\exp(-t^2/2T_0^2)+\sqrt{\xi}\exp(-t^2/2T_0^2-i\Delta\omega t)]$, is the
superposition of a pump pulse and a weak signal, whose frequency is shifted by $\Delta\omega$. The
ratio $\xi=P_{s}/P_{p}$ is set to \SI{10}{\percent} and \SI{1}{\percent} in the cases of
wavelength conversion and parametric amplification, respectively, so that in the latter case the
signal is too weak to affect the pump. We also assume that the signal and pump have the same
temporal width, $T_{0}$, and, unless otherwise stated, the same group-velocity, $v_{g}$.

\begin{figure}[htb]
\centerline{\includegraphics[width=8cm]{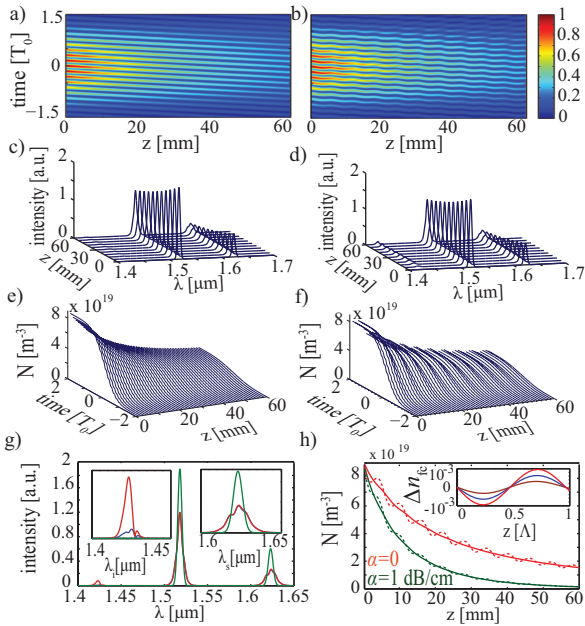}} \caption{Left (right) panels show the
evolution of an optical pulse in a uniform (quasi-phase-matched Bragg) waveguide (see the text for
the values of the pulse and waveguide parameters). Top, second, and third row panels show the
$z$-dependence of the temporal pulse profile, its spectrum, and FC density, respectively. (g)
Input (green) and output pulse spectra corresponding to the uniform (blue) and Bragg (red)
waveguides. In inset, the signal and pump regions of the spectra. (h) Variation $N(z)$, for
uniform (---) and Bragg ($\cdots$) waveguides. In inset, dependence $\Delta n_{\mathrm{fc}}(z)$,
for $\Delta w=$~\SI{10}{\nano\meter} (brown), $\Delta w=$~\SI{20}{\nano\meter} (blue), and $\Delta
w=$~\SI{30}{\nano\meter} (red).} \label{const_modul}
\end{figure}

A generic example of pulse evolution in a uniform and Bragg Si-PNW, where the latter is designed
such that condition \eqref{Dbeta} holds, is presented in Fig.~\ref{const_modul}. We considered a
pulse with $T_{0}=$~\SI{500}{\femto\second}, $P_p=$~\SI{200}{\milli\watt},
$P_s=$~\SI{20}{\milli\watt}, $\lambda_p=$~\SI{1518}{\nano\meter}, and
$\lambda_s=$~\SI{1623}{\nano\meter}, so that one expects an idler pulse to form at
$\lambda_i=$~\SI{1426}{\nano\meter}. The waveguide parameters are $w_0=$~\SI{740}{\nano\meter},
$\Delta w=$~\SI{30}{\nano\meter}, $\Lambda=$~\SI{6}{\milli\meter},
$\beta_{2,p}=$~\SI{0.15}{\pico\second\squared\per\meter},
$\beta_{4,p}=$~\SI{-6.1e-7}{\pico\second\tothe{4}\per\meter}, and
$\gamma^{\prime}_{p}=$~\SI{201.4}{\per\watt\per\meter}. The evolution of the temporal pulse
profile, shown in Figs.~\ref{const_modul}(a) and \ref{const_modul}(b), suggests that the pulse
propagates with a group-velocity, $v_{g}$, slightly larger than $v_{g}(\omega_{p})$. Indeed, the
pulse propagates in the normal dispersion regime and its average frequency is smaller than
$\omega_{p}$, which means that $v_{g}>v_{g}(\omega_{p})$. In the case of the Bragg waveguide,
additional temporal oscillations of the pulse are observed. This effect is traced to the periodic
variation $v_{g}(z)$, which is due to the implicit dependence of $v_{g}$ on a periodically varying
width $w(z)$.

Due to its specific nature, it is more suitable to study the FWM in the frequency domain. In
particular, the differences between the evolution of the pulse spectra in uniform and Bragg
waveguides, illustrated by Figs.~\ref{const_modul}(c) and \ref{const_modul}(d), respectively,
underline the main physics of pulsed FWM in Si-PNWs. Specifically, it can be seen that, in the
Bragg waveguide, the idler energy builds up at a much higher rate as compared to the case of the
uniform Si-PNW, an indication of a much more efficient FWM interaction [see also
Fig.~\ref{const_modul}(g)]. In both cases, however, we observe a gradual decrease of the the pulse
peak power, induced by the linear and nonlinear losses associated to the generated FCs. Note that
the dispersion length $L_{d}=T_{0}^{2}/\vert\beta_{2}\vert\approx$~\SI{1.6}{\meter} so that the
dispersion-induced pulse broadening is negligible.

For the Bragg waveguide one can also observe a series of oscillations of the FC density with
respect to $z$, which are due to the periodic variation with $z$ of $\gamma^{\prime\prime}$.
Specifically, the oscillatory $z$-variation of $N(z)$ results in a quasi-periodic variation of the
effective modal index, $\Delta n_{\mathrm{fc}}(z)$, which adds to the periodic variation of
$n_{\mathrm{eff}}$ due to the waveguide-width modulation. Note, however, that for the power values
used in this analysis the former effect is an order of magnitude weaker than the latter one
[compare Fig.~\ref{Pha_match}(c) with the inset in Fig.~\ref{const_modul}(h)].

\begin{figure}[htb]
\centerline{\includegraphics[width=8cm]{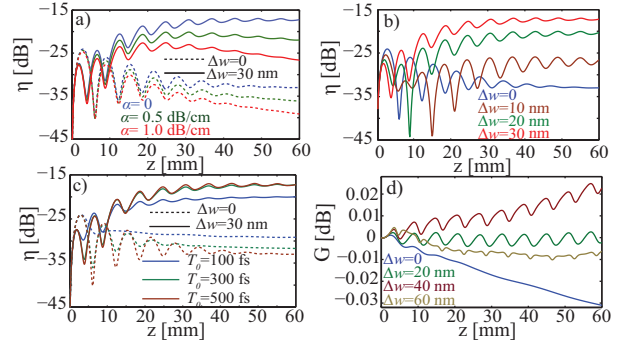}} \caption{(a), (b) CE $\eta(z)$, calculated
for different $\alpha$ and $\Delta w$, respectively. $\alpha=0$ in (b). (c) CE calculated for
different $T_{0}$, for Bragg (---) and uniform ($\cdots$) waveguides. Pulse and waveguide
parameters in (a)--(c) are the same as in Fig.~\ref{const_modul}. (d) FWM gain vs. $z$ (the values
of pulse and waveguide parameters are given in the text).} \label{CE_Gain}
\end{figure}
A comparative study of the conversion efficiency (CE), $\eta(z)=10\log [E_{i}(z)/E_{s}(0)]$, and
FWM gain, $G(z)=E_{s}(z)/E_{s}(0)$, in a Bragg vs. a uniform Si-PNWs is summarized in
Fig.~\ref{CE_Gain}. The energies of the idler, $E_{i}$, and signal, $E_{s}$, were calculated by
integrating the power spectrum over a frequency domain containing the corresponding pulse. These
results clearly show that the Bragg grating induces a dramatic increase of the CE. Although the CE
decreases with the waveguide loss, the CE enhancement between the uniform and Bragg waveguides
only slightly varies with $\alpha$. Importantly, the power decay leads to the detuning of the FWM
and, after a certain distance, to the degradation of its efficiency. As expected, the CE
enhancement increases with $\Delta w$, reaching \SI{15}{\decibel} for $\Delta
w=$~\SI{30}{\nano\meter}. The CE also depends on $T_{0}$, as per Fig.~\ref{CE_Gain}(c). Indeed,
one expects that the CE increases with $T_{0}$ since the Bragg waveguide is designed to
phase-match the carrier frequencies of the pulses, so that spectrally narrower pulses are better
phase-matched.

The dependence of the FWM gain on the amplitude of the width modulation is shown in
Fig.~\ref{CE_Gain}(d). To avoid large losses due to two-photon absorption, the device is operated
at mid-IR frequencies. Thus, the pulse has $T_{0}=$~\SI{500}{\femto\second},
$P_p=$~\SI{200}{\milli\watt}, $P_s=$~\SI{2}{\milli\watt}, $\lambda_p=$~\SI{2215}{\nano\meter}, and
$\lambda_s=$~\SI{2102}{\nano\meter}, meaning that the idler is formed at
$\lambda_i=$~\SI{2340}{\nano\meter}. The waveguide parameters were $w_0=$~\SI{720}{\nano\meter},
$\Lambda=$~\SI{6}{\milli\meter}, $\beta_{2,p}=$~\SI{0.43}{\pico\second\squared\per\meter},
$\beta_{4,p}=$~\SI{3.4e-4}{\pico\second\tothe{4}\per\meter}, and
$\gamma^{\prime}_{p}=$~\SI{92.8}{\per\watt\per\meter}. The increased FWM efficiency in Bragg
Si-PNWs is clearly demonstrated by these numerical experiments namely, a transition from negative
to positive net gain is observed when $\Delta w$ increases from zero to \SI{40}{\nano\meter}. When
$\Delta w$ further increases beyond a certain value, $\Delta w\approx$~\SI{50}{\nano\meter}, the
variation over one period of $\beta$ becomes large enough to greatly degrade the phase matching of
the interacting pulses, resulting in a steep decrease of the FWM gain.

\begin{figure}[htb]
\centerline{\includegraphics[width=8cm]{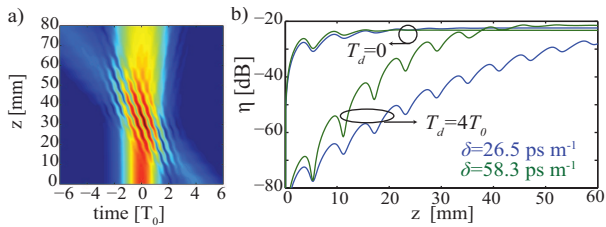}} \caption{(a) Pulse evolution for
$\lambda_s=$~\SI{2181}{\nano\meter} and $\lambda_i=$~\SI{2066}{\nano\meter}. (b) CE dependence on
$z$. Green and blue lines correspond to the pulse in (a) and $\lambda_s=$~\SI{2066}{\nano\meter}
and $\lambda_i=$~\SI{2181}{\nano\meter}, respectively. The other parameters in (a) and (b) are
$w_0=$~\SI{600}{\nano\meter}, $\Lambda=$~\SI{6}{\milli\meter},
$\lambda_p=$~\SI{2122}{\nano\meter}, $T_{0}=$~\SI{500}{\femto\second},
$P_p=$~\SI{200}{\milli\watt}, and $P_s=$~\SI{20}{\milli\watt}.} \label{walkoff}
\end{figure}

In our analysis so far we have designed the waveguide so that the pump and signal have the same
group-velocity, meaning that optimum FWM is then achieved. In Fig.~\ref{walkoff}, which also
considers mid-IR pulses, we present the CE determined in two cases when this condition does not
hold, i.e. when the walk-off $\delta=\vert 1/v_{g,p}-1/v_{g,s}\vert\neq0$, and for two different
values of the pump-signal time delay, $T_{d}$. The main conclusion that can be drawn from these
results is that when $\delta\neq0$, FWM occurs only over a certain distance, which is related to
the time necessary for the pump and signal pulses to pass through each other. In
Fig.~\ref{walkoff}(a) this propagation section corresponds to the region where one can observe a
series of intensity fringes, which are due to the frequency beating between the two pulses. Also,
the CE increases rapidly as $T_{d}$ decreases because for large $T_{d}$ the pump decays more
before it begins to interact with the signal, i.e. the FWM becomes more detuned. This suggests
that the CE should increase with $\delta$ as well, in agreement with the results plotted in
Fig.~\ref{walkoff}(b) for $T_{d}=4T_{0}$.

In conclusion, we showed that efficient pulsed FWM can be achieved in long-period Bragg silicon
waveguides, which can be used for pulse amplification and to enhance the wavelength-conversion
efficiency, as compared to uniform waveguides. These new ideas can be applied to a multitude of
photonic devices, including photonic crystal fibers and sub-micrometer optical waveguides whose
modal frequency dispersion is primarily determined by the waveguide dispersion. Equally important,
by using more complex grating profiles, e.g. multi-period \cite{dog12oe} or chirped gratings
\cite{ncs08prl,yf10pra}, one can design photonic devices with enhanced functionality, including
ultra-broadband sources of entangled photons and highly efficient autoresonant optical parametric
amplifiers.

The work of S. L. was supported through a UCL Impact Award. R. R. G. acknowledges support from the
Columbia Optics and Quantum Electronics IGERT.


\begin{thebibliography}{99}

\bibitem{sba74apl} R. H. Stolen, J. E. Bjorkholm, and A. Ashkin, \apl \textbf{24}, 308 (1974).

\bibitem{hjk78jap} K. O. Hill, D. C. Johnson, B. S. Kawasaki, and R. I. MacDonald, \jap
\textbf{49}, 5098 (1978).

\bibitem{crd03oe} R. Claps, V. Raghunathan, D. Dimitropoulos, and B. Jalali, \opex \textbf{11},
2862 (2003).

\bibitem{drc04oe} D. Dimitropoulos, V. Raghunathan, R. Claps, and B. Jalali, \opex \textbf{12},
149 (2004).

\bibitem{fys05oe} H. Fukuda, K. Yamada, T. Shoji, M. Takahashi, T. Tsuchizawa, T. Watanabe, J.
Takahashi, and S. Itabashi, \opex \textbf{13}, 4629 (2005).

\bibitem{edo05oe} R. Espinola, J. Dadap, R. M. Osgood, Jr., S. McNab, and Y. Vlasov, \opex
\textbf{13}, 4341 (2005).

\bibitem{fts06n} M. A. Foster, A. C. Turner, J. E. Sharping, B. S. Schmidt, M. Lipson, and A. L.
Gaeta, \nat \textbf{441}, 960 (2006).

\bibitem{lzf06oe} Q. Lin, J. Zhang, P. M. Fauchet, and G. P. Agrawal, \opex \textbf{14}, 4786
(2006).

\bibitem{pco06ol} N. C. Panoiu, X. Chen, and R. M. Osgood, \ol \textbf{31}, 3609 (2006).

\bibitem{yft06ptl} K. Yamada, H. Fukuda, T. Tsuchizawa, T. Watanabe, T. Shoji, and S. Itabashi, \ptl
\textbf{18}, 1046 (2006).

\bibitem{krs06oe} Y.-H. Kuo, H. Rong, V. Sih, S. Xu, M. Paniccia, and O. Cohen, \opex \textbf{14},
11721 (2006).

\bibitem{fts07oe} M. A. Foster, A. C. Turner, R. Salem, M. Lipson, and A. L. Gaeta, \opex
\textbf{15}, 12949 (2007).

\bibitem{lov10np} X. Liu, R. M. Osgood, Y. A. Vlasov, and W. M. J. Green, \np \textbf{4}, 557 (2010).

\bibitem{zpm10np} S. Zlatanovic, J. S. Park, S. Moro, J. M. C. Boggio, I. B. Divliansky, N. Alic,
S. Mookherjea, and S. Radic, \np \textbf{4}, 561 (2010).

\bibitem{klo11oe} B. Kuyken, X. Liu, R. M. Osgood, R. Baets, G. Roelkens, and W. Green,
Opt. Exp. \textbf{19}, 20172 (2011).

\bibitem{lkr12np} X. Liu, B. Kuyken, G. Roelkens, R. Baets, R. M. Osgood, and W. M. J. Green, \np
\textbf{6}, 667 (2012).

\bibitem{dog12oe} J. B. Driscoll, N. Ophir, R. R. Grote, J. I. Dadap, N. C. Panoiu, K. Bergman,
and R. M. Osgood, \opex \textbf{20}, 9227 (2012).

\bibitem{lpa07oe} Q. Lin, O. J. Painter, and G. P. Agrawal, \opex \textbf{15}, 16604 (2007).

\bibitem{opd09aop} R. M. Osgood, N. C. Panoiu, J. I. Dadap, X. Liu, X. Chen, I-W. Hsieh, E.
Dulkeith, W. M. J. Green, and Y. A. Vlassov, \aop \textbf{1}, 162 (2009).

\bibitem{la06ol} Q. Lin, and G. P. Agrawal, \ol \textbf{31}, 3140 (2006).

\bibitem{sft08np} R. Salem, M. A. Foster, A. C. Turner, D. F. Geraghty, M. Lipson, and A. L.
Gaeta, \np \textbf{2}, 35 (2008).

\bibitem{cpo06jqe} X. Chen, N. C. Panoiu, and R. M. Osgood, \jqe \textbf{42}, 160 (2006).

\bibitem{plo09ol} N. C. Panoiu, X. Liu, and R. M. Osgood, \ol \textbf{34}, 947 (2009).

\bibitem{ldj13ol} S. Lavdas, J. B. Driscoll, H. Jiang, R. R. Grote, R. M. Osgood, and N. C.
Panoiu, \ol \textbf{38}, 3953 (2013).

\bibitem{sb87jqe} R. A. Soref and B. R. Bennett, \jqe \textbf{23}, 123 (1987).

\bibitem{ncs08prl} M. B. Nasr, S. Carrasco, B. E. A. Saleh, A. V. Sergienko, M. C. Teich, J. P.
Torres, L. Torner, D. S. Hum, and M. M. Fejer, \prl \textbf{100}, 183601 (2008).

\bibitem{yf10pra} O. Yaakobi and L. Friedland, \pra \textbf{82}, 023820 (2010).

\end{thebibliography}
\end{document}